\documentclass[10pt,aps,prd,nofootinbib,superscriptaddress,twocolumn]{revtex4}
\usepackage[utf8]{inputenc}
\DeclareUnicodeCharacter{200B}{{\hskip 0pt}}
\usepackage{color}
\usepackage{graphicx}
\usepackage{amsmath}
\usepackage{amssymb}
\usepackage{bm}
\usepackage{acronym}
\usepackage{ifthen}
\usepackage{blindtext}
\usepackage[normalem]{ulem}
\usepackage{hyperref}
\usepackage{etoolbox}
\usepackage{fancyhdr}
\usepackage{xspace}
\usepackage{textcomp}
\usepackage{multirow}
\usepackage[caption=false]{subfig}
\usepackage{lineno}
\usepackage{tabularx}
\hypersetup{
    colorlinks=true,
    linkcolor=blue,
    filecolor=magenta,      
    urlcolor=black,
		citecolor=red,
		}
\begin{document}

\title{Gravitational Energy Creation in the Sandwich pp-Waves Collision}

\author{F. L. Carneiro}\thanks{Corresponding author: fernandolessa45@gmail.com}

\affiliation{Universidade Federal do Norte do Tocantins, 77824-838, Aragua\'ina, TO, Brazil}

\author{K. Q. Abbasi}
\email{kamran.qadir@sns.nust.edu.pk}
\affiliation{Department of Mathematics,
Faculty of Engineering and Computing $($FE\&C$)$, National University of Modern Languages $($NUML$)$, Sector, H-9, Islamabad, Pakistan.}
\affiliation{Department of Mathematics,
School of Natural Sciences $($SNS$)$, National University of Sciences and Technology $($NUST$)$, Sector, H-12, Islamabad, Pakistan.}

\begin{abstract}
This article investigates the spacetime of two colliding sandwich gravitational waves, focusing on evaluating gravitational energy before and after the collision. In the framework of the Teleparallel Equivalent of General Relativity (TEGR), we derive a true energy-momentum tensor for the gravitational waves and integrate it over a finite region of space, obtaining analytical expressions for the energy of each wave and the resulting spacetime. Our findings reveal that the energy after the collision exceeds the pre-collision, indicating energy creation. We analyze the energy density and ``surface energy density" on the wavefronts, underscoring their divergence near the singularity. Additionally, we observe that the colliding waves drag observers but exert no acceleration at the collision event. This study addresses and resolves longstanding issues raised by Szekeres in his seminal work on colliding pp-waves, offering a more physically realistic framework through the local energy definition provided by TEGR. The implications for gravitational wave interactions and their energy transfer mechanisms are discussed.
\end{abstract}


\maketitle

\date{\today}

\section{Introduction}

The development of Einstein's General Relativity (GR) in 1915 drew significant attention from the scientific community toward the pursuit of exact solutions for the theory's field equations. Although consistent---comprising ten equations and ten unknown components of the fundamental variable, the metric tensor---the non-linearity of these equations makes finding analytical solutions challenging. However, symmetry considerations can simplify the problem and enable the identification of exact classes of solutions.

In the early days of GR, Einstein and Rosen investigated the electromagnetic plane waves of Maxwell Electrodynamics (ME) as an analogy to search for gravitational wave solutions in GR. During the 1920s and 1930s, they explored these solutions but initially dismissed them as nonphysical, as they required a single nonsingular coordinate system to cover the entire spacetime. It was only in the late 1950s that Bondi and Pirani established gravitational waves as a valid physical class of solutions to Einstein's equations \cite{bondi1957plane,bondi1959gravitational}.

Plane gravitational waves are defined as a congruence of null geodesics in spacetime, analogous to the congruence of null rays of plane electromagnetic waves. Since the wavefronts are two-dimensional plane surfaces and parallel to each other, the separation vector of the normal null 4-vectors must remain constant. In other words, the congruence is characterized by the absence of expansion, distortion, or rotation. Consequently, plane gravitational waves are referred to as plane-fronted gravitational waves with parallel rays (pp-waves).

Much attention has been given to pp-waves in recent decades, as they are exact and nonlinear solutions to Einstein's equations, in contrast to linearized gravitational waves that arise as solutions of the linearized Einstein equations. One of the most intriguing features of gravitational waves is the memory effect, i.e., a pulse of gravitational waves can permanently alter the configuration of particles, their separation vector after the pulse differs from before its passage. Interestingly, memory effects are associated with the presence of Burgers vectors \cite{abid2024identification}, which measure topological dislocation defects in matter.

Memory effect is a non-linear phenomenon, challenging to track in numerical simulations. It may manifest when a linear gravitational wave sources another wave \cite{favata2010gravitational}, a consequence of the non-linearity of the GR field equations, which allow gravitational fields to act as sources for new fields. Since pp-waves are inherently non-linear, they can describe the memory effect through numerical solutions of the geodesic equations \cite{zhang2017memory}, where the pulse is modulated as a Gaussian.

Recently, the geodesic equations were studied for a ramp profile and analytically solved in Brinkmann coordinates. The memory effect was also observed in this case, as it has been for several other pulse profiles, such as square \cite{chakraborty2022simple} and impulse \cite{zhang2018memory} pulses, further corroborating the universality of this effect for pp-waves \cite{datta2024memory}.  
Furthermore, recent field-theoretic computations have explored the gravitational wave memory effect in eccentric and hyperbolic binary orbits, providing waveform templates in the frequency domain that facilitate signal extraction from observational data \cite{hait2024frequency}.

The memory effect has also been studied within Einstein-Cartan theory \cite{cvetkovic2022memory}, where the role of torsion in spacetime geometry was considered, revealing a connection to the mass of the tordion---a particle-like excitation of torsion fields. Although difficult to detect using Earth-based interferometers, there is potential for detecting the memory effect with space-based interferometers, such as the future TianQin mission \cite{sun2023detecting}.
Additionally, investigations into the gravitational wave memory signal produced by neutrinos in supernovae suggest that self-interacting neutrinos generate weaker high-frequency memory signals compared to free-streaming neutrinos, with implications for their detection by future observatories like DECIGO and BBO \cite{bhattacharya2024gravitational}.

Later, it was discovered that non-linear effects also influence the velocity of particles \cite{zhang2018velocity,divakarla2021first}, i.e., a pp-wave may permanently alter a particle's velocity. Although there are specific parameter values ​​that do not produce velocity memory effects \cite{zhang2024displacement,zhang2024gravitational}.

The velocity memory effect has significant implications for our understanding of pp-waves. When dealing with a finite pulse, spacetime is mostly flat before and after the wave's passage. Thus, in the regime of low velocities, classical mechanics is expected to apply. If the wave changes the particle's velocity, its kinetic energy is also altered. Since the wave is the only entity interacting with the particle, it can be concluded that this energy alteration results from an exchange of energy between the particle and the wave. Depending on the initial conditions, the particle may gain or lose energy, implying that a pp-wave can either provide or remove energy from a particle \cite{maluf2018plane}.

This randomness in the energy transfer between a pp-wave and a particle may help explain how pp-waves propagate over long distances in the universe without dissipating when interacting with matter. If they persist for vast periods, pp-waves are expected to collide with one another frequently throughout the universe.

Besides the initial analogy between pp-waves and electromagnetic waves, when two electromagnetic waves interact in vacuum, they locally superpose and later continue their paths without retaining a ``memory" of the interaction. However, since ME is a linear theory, the non-linear features of GR must be considered when dealing with colliding pp-waves.

In 1970, Szekeres studied two plane gravitational waves with constant $+$ polarization \cite{szekeres1970colliding}, whose line element in Brinkmann coordinates is given by
\begin{equation}\label{eq1}
ds^{2}=dx^{2}+dy^{2}-2dudv + f(u)(x^{2}-y^{2})du^{2}\,, 
\end{equation}
where the spatial coordinates ${x,y}$ span the two-dimensional flat wavefronts, ${u,v}$ are null coordinates, and $f(u)$ describes the wave pulse. Although Brinkmann coordinates are more general, allowing the description of more polarization states, Szekeres resorted to transforming (\ref{eq1}) into Rosen coordinates, where the $+$ polarization is described as
\begin{equation} 
ds^{2}=F(u)^{2}dx^{2}+G(u)^{2}dy^{2}-2e^{M(u)}dudv\,.
\end{equation}

Considering two sandwich pp-waves---characterized by a region of non-vanishing curvature tensor surrounded by two flat regions---moving in opposite directions, Szekeres divided spacetime into regions before and after the collision. He used the known solutions in the pre-collision regions and boundary conditions to solve the initial value problem for the evolution equations in the collision region. He was able to find an exact class of solutions, the aspects of which are discussed in detail in Section \ref{sec2}.

Szekeres found that the coordinate system becomes singular in the collision region, regardless of the wave strength, and that the Weyl invariant---constructed from the curvature tensor projected into null tetrads---also becomes singular. Thus, the singularity is not merely a coordinate artifact but a physical one. Szekeres considered this problem less significant, as real gravitational waves would not have perfectly plane surfaces but rather curved ones, and he speculated that curved wavefronts might prevent the appearance of singularities.

Later, in 1972, while revising his results, Szekeres examined the energy of the incoming gravitational waves and the energy post-collision \cite{szekeres1972colliding}. It is well known that GR lacks a proper definition for the energy-momentum tensor of the gravitational field, a topic we shall address shortly. Thus, Szekeres employed the Landau-Lifshitz pseudotensor under a linear approximation. He found an infinite energy content for the incoming waves and an indefinitely long energy flux after the collision. Additionally, he considered two definitions of energy flux, yielding results of infinite and zero flux, respectively.  

A plane gravitational wave is an idealization of a real gravitational wave far from its source. Since the wave disperses over an infinite wavefront, it is expected that the total energy---calculated by integrating the pseudotensor over the entire space---results in an infinite value. Therefore, physical interpretation is limited to the energy density, which can be integrated over a localized region of space. The issue is that pseudotensors are coordinate-dependent, meaning the energy calculation varies with the coordinate system, violating the Principle of General Covariance. Consequently, defining a true energy-momentum tensor for the gravitational field is necessary to resolve this problem.  

The search for a true energy-momentum tensor paralleled the development of GR, culminating in M{\o}ller's conclusion in the 1960s that no such definition was possible if the metric tensor $g_{\mu\nu}$ was treated as the fundamental variable of gravitation \cite{moller1964conservation}. M{\o}ller argued that tetrads $e_{a\mu}$ must be regarded as the fundamental variables to define a proper energy-momentum tensor and localize gravitational energy. Following this idea, Maluf in the 1990s solved the energy localization problem \cite{maluf1995localization}, employing a tetrad formulation of gravitation that leads to field equations equivalent to Einstein's, i.e., the TEGR.  

TEGR, originally considered by Einstein in an attempt to unify electromagnetism and gravitation, provided the perfect framework for defining the local energy of the gravitational field. While some argue that defining local energy conflicts with the equivalence principle, this does not preclude the existence of a well-defined energy for the gravitational field \cite{formiga2023gravitational}. The TEGR energy expression has been applied to calculate the energy of various known solutions to Einstein's equations, including regular pp-waves in Brinkmann coordinates \cite{maluf2008energy}, where an integral expression was obtained.

In this article, we consider the spacetime of two colliding pp-waves: one propagating in the positive  $u$ direction and the other in the positive $v$ direction. Our aim is to address the initial problem identified by Szekeres, namely, the evaluation of the energy of the spacetime before and after the collision of the waves. Using the local definition of energy from TEGR, we calculate the energy-momentum tensor of the gravitational field and integrate it over a finite section of space, obtaining an analytical expression for the energy of each wave as well as for the spacetime resulting from their collision.

The local definition of energy is of key importance, as it renders a physically realistic result. The energy of an idealized plane wave may be considered an approximation of a sufficiently small section of a curved wavefront, which can be realistically approximated by a plane front within an appropriately defined area. Therefore, the local result provides a good approximation of a realistic case.

We first investigate the energy density itself, as we have a true energy-momentum tensor that covariantly transforms under coordinate transformations. Initially, we examine stationary observers in the Szekeres spacetime and find that the incoming waves drag observers but exert no acceleration at the collision event. Furthermore, we observe that it is impossible for an observer to remain stationary near the singularity. We obtain a ``surface energy density," which is a function of the null coordinates $ u $ and $ v $. We find that one observer measures a positive value for one wave and a negative value for the other. Consequently, the surface energy density is zero at the collision event but assumes positive and negative values as we approach the singularity, ultimately diverging at the singularity.

By converting the double null coordinates into Cartesian-like coordinates, $ t = \frac{u+v}{\sqrt{2}} $ and $ z = \frac{u-v}{\sqrt{2}} $, we integrate the expressions over a finite region of space and track the variation of energy with time. Since we have a local expression for the energy, we evaluate the energy of the two incoming waves separately before the collision, i.e., in two distinct regions: for the first wave and for the second wave. We then compare their sum with the energy in the whole region after the collision. We find that the energy after the collision is greater than before. Ergo, energy is created following the collision.

This article is divided as follows. In section \ref{sec2}, we briefly present TEGR and highlight the main results necessary for the comprehension of our analysis. In section \ref{sec3}, we discuss some mathematical aspects of the Szekeres spacetime and explain the regions and solutions that will be considered in the next section. In section \ref{sec4}, we evaluate the expressions for the gravitational energy density presented in section \ref{sec2} for the spacetime considered in section \ref{sec3}. In section \ref{sec5}, we evaluate the energy of the spacetime resulting from the collision. Finally, in section \ref{conc}, we present our conclusions.

In order to distinguish between the local and spacetime indices of the tetrads, we use the following notation. Spacetime indices $\mu,\nu,\ldots$ are denoted by $0,1,2,3$; Lorentz indices $a,b,\ldots$ are denoted by $(0),(1),(2),(3)$. The spacetime metric tensor $g_{\mu\nu}$ raises and lowers spacetime indices, and the flat spacetime metric tensor $\eta_{ab}$ raises and lowers the Lorentz indices. We use a signature $(-,+,+,+)$ and the geometrized unit system, where $c=G=1$.

\section{Teleparallel Equivalent to General Relativity}\label{sec2}

TEGR provides a geometrical description of gravity, treating the tetrad field $ e^{a}\,_{\mu} $ as the fundamental variable of the gravitational field. Through the relation $ e^{a\mu} = g^{\mu\nu}e^{a}\,_{\nu} $, the metric tensor can be expressed as  
\begin{equation}\label{eq3}
	g_{\mu\nu}=e^{a}\,_{\mu}e_{a\nu}\,,
\end{equation}
demonstrating that tetrads are more fundamental entities, encapsulating the geometry of spacetime. Consequently, a theory that is a function of the metric is also a function of the tetrads.  

The tetrads possess sixteen independent components, associated with two distinct indices. In order to elucidate the physical significance of the additional components, we examine their transformation properties. The Greek index transforms as a 4-vector under coordinate transformations, while the Latin index transforms as a 4-vector under Lorentz transformations $ \Lambda^{a}\,_{b} $, i.e.,
$
	\tilde{e'}^{a}\,_{\mu}=\Lambda^{a}\,_{b}\frac{\partial x ^\nu}{\partial x'^{\mu}}e^{b}\,_{\nu}\,,
$
where tildes indicate frame transformations and primes denote coordinate transformations. The set of four linearly independent vectors, $ \{e_{(0)}\,^{\mu}, e_{(i)}\,^{\mu}\} $, defines the local coordinate system of an observer moving along their worldline in spacetime. Thus, the Latin indices $ a $ represent the flat tangent spacetime at the event $ x^{\mu}(\tau) $, where $ \tau $ is the observer's proper time. Consequently, tetrads establish the Inertial Frame of Reference (IFR) for the observer.  

Since the observer is always at rest in their own IRF, their worldline can only be oriented along the $ e_{(0)}\,^{\mu} $ direction, i.e., $e_{(0)}\,^{\mu} $ is tangent to the observer's worldline and can be related to their 4-velocity $ u^{\mu} $, i.e., $ e_{(0)}\,^{\mu} = u^{\mu} $.
Hence, the observer’s 4-acceleration can be expressed as  
\begin{equation}\label{eq4}
	\frac{Du^{\mu}}{d\tau}=u^{\nu}\partial_{\nu}e_{(0)}\,^{\mu}+\mathring{\Gamma}^{\mu}\,_{\nu\lambda} u^{\nu}e_{(0)}\,^{\lambda}\,,
\end{equation} 
where $ \mathring{\Gamma}^{\mu}\,_{\nu\lambda} $ are the Christoffel symbols. From this, we observe that the tetrads not only carry information about spacetime geometry but also about the kinematic state of the observer. Therefore, from the covariant derivative of the tetrads, we can define the antisymmetric acceleration tensor $ \phi_{ab} $ as  
\begin{equation}
	\frac{De_{a}\,^{\mu}}{d\tau}=\phi_{a}\,^{b}e_{b}\,^{\mu}.
\end{equation}
Since $ \phi_{(0)}\,^{(i)} $ yields zero when a 4-acceleration is absent, we identify the three components $ \phi_{(0)}\,^{(i)} $ as the inertial acceleration of the frame in the $ (i) $-direction. The remaining components $ \phi_{(i)}\,^{(j)} $ represent the rotational frequency of the frame relative to a non-rotating Fermi-Walker transported frame \cite{mashhoon2002length,mashhoon2003vacuum}. The acceleration tensor, therefore, characterizes the kinematic state of the observer.

\subsection{Weitzenb\"ock geometry}

The geometry of a metric-affine spacetime is determined by the properties of the affine connection. A geometry may have, independently, torsion, curvature, and non-metricity. If we assume the geometry to have metricity, i.e., $ \nabla_{\lambda}g_{\mu\nu} = 0 $, we may have the torsion 2-form
\begin{equation}\label{eq6}
	T^{a}=\frac{1}{2}\Big(\partial_{\mu}e^a\,_{\nu}-\partial_{\nu}e^a\,_{\mu}+\omega^{a}\,_{b\mu}e^{b}\,_{\nu}-\omega^{a}\,_{b\nu}e^{b}\,_{\mu}\Big)dx^{\mu}\wedge dx^{\nu}\,,
\end{equation}
and the curvature 2-form
\begin{align}
	R^{a}\,_{b}
	&=\frac{1}{2}\Big( \partial_{\mu}\omega^{a}\,_{b\nu} - \partial_{\nu}\omega^{a}\,_{b\mu}\nonumber\\
	& \quad+ \omega^{a}\,_{c\mu}\omega^{c}\,_{b\nu} - \omega^{a}\,_{c\nu}\omega^{c}\,_{b\mu}
\Big)dx^{\mu}\wedge dx^{\nu}\,,\label{eq7}
\end{align}
where ``$\wedge$" denotes the exterior product and $ \omega^{a}\,_{b\mu} $ are the 0-form components of the 1-form connection $\omega^{a}\,_{b}=\omega^{a}\,_{b\mu}dx^{\mu}$.

In a geometry where the tetrads are parallel transported along a closed finite path, there exists a connection where the covariant derivative of the tetrads vanishes, i.e.,
\begin{equation}
	\nabla_{\mu}e^{a}\,_{\nu}=\partial_{\mu}e^{a}\,_{\nu} -\Gamma^{\lambda}\,_{\mu\nu}e^{a}\,_{\lambda}=0\,.
\end{equation}
The connection that guarantees the parallelism $ \Gamma^{\lambda}\,_{\mu\nu} $ is the Weitzenb\"ock connection. This connection yields a non-zero torsion tensor 0-form components
\begin{equation}\label{eq9}
	T^{\lambda}\,_{\mu\nu}=\Gamma^{\lambda}\,_{\mu\nu}-\Gamma^{\lambda}\,_{\nu\mu}=e_{a}\,^{\lambda}\partial_{\mu}e^{a}\,_{\nu}-e_{a}\,^{\lambda}\partial_{\nu}e^{a}\,_{\mu}\,.
\end{equation}
TEGR is a gravitational description of gravity where spacetime geometry is the Weitzenb\"ock one. In TEGR, the affine connection $ \omega^{a}\,_{b\mu} $ plays no role in the dynamics \cite{maluf2013teleparallel}, thus it can be considered zero. Cartan structure equations (\ref{eq6}) and (\ref{eq7}) then reduce to
\begin{equation}
	T^{a}=\frac{1}{2}\Big(\partial_{\mu}e^{a}\,_{\nu}-\partial_{\nu}e^{a}\,_{\mu}\Big)dx^{\mu}\wedge dx^{\nu}\,,
\end{equation}
and $R^{a}\,_{b}=0$.

Given the nullity of the connection $ \omega^{a}\,_{b\mu} $, the components of the Levi-Civita connection $ \mathring{\omega}_{\mu a b} $ are minus those of the contortion tensor $ K_{\mu ab} $, i.e.,
\begin{equation}\label{eq11}
	-\mathring{\omega}_{a b\mu}=K_{ ab\mu}=\frac{1}{2}e_{a}\,^{\alpha}e_{b}\,^{\beta}\Big( T_{\mu\alpha\beta} + T_{\alpha\mu\beta}- T_{\beta\mu\alpha}\Big)\,.
\end{equation}
From the above relation, the acceleration tensor may be written as \cite{maluf2013teleparallel}
\begin{equation}\label{eq12}
\phi_{ab}=\frac{1}{2}\Big( T_{(0)ab} + T_{a(0)b} - T_{b(0)a} \Big)\,.
\end{equation}
Thus, the torsion tensor allows the description of the inertial state of an observer in spacetime.

\subsection{Field Equations}

From equation (\ref{eq11}), we can see that the curvature scalar constructed from the curvature tensor of the Levi-Civita torsion-free connection is a function of the contortion and, as a consequence, of the tetrads. It yields, after some lengthy calculations,
\begin{equation}\label{eq13}
	eR(e)=-e\Sigma^{abc}T_{abc}+2\partial_{\mu}(eT^{\mu})\,,
\end{equation}
where $ e $ is the determinant of the tetrads, $ k = \frac{1}{16\pi} $ is the coupling constant, $ T^{\mu} = T^{\lambda}\,_{\lambda}\,^{\mu} $ is the trace of the torsion tensor, and the superpotential is given by
\begin{equation}\label{eq14}
\Sigma^{abc}=\frac{1}{4} (T^{abc}+T^{bac}-T^{cab}) +\frac{1}{2}(
\eta^{ac}T^b-\eta^{ab}T^c)\,.
\end{equation}
If we construct the Lagrangian density from the scalar (\ref{eq13}), we obtain the Einstein-Hilbert action, which would yield a Lagrangian density invariant under coordinate transformations and local and global Lorentz transformations. The invariance under local Lorentz transformations is contained in the divergent term in (\ref{eq13}). This term does not affect the field equations. Hence, the TEGR Lagrangian density is defined as
\begin{equation}\label{eq15}
\mathcal{L}=-ke\Sigma^{abc}T_{abc}-\mathcal{L}_{M}\,,
\end{equation}
where $ \mathcal{L}_{M} $ stands for the Lagrangian density of the matter-radiation fields. By constructing the action from (\ref{eq15}) and varying it with respect to $ e_{a\mu} $, we obtain \cite{maluf2013teleparallel}
\begin{equation}\label{eq16}
\partial_{\nu}\left(e\,\Sigma^{a\lambda\nu}\right)=\frac{1}{4k}e\,e^{a}\,_{\mu}\left(t^{\lambda\mu}+T^{\lambda\mu}\right)\,,
\end{equation}
where $ T^{\lambda\mu} $ is the energy-momentum tensor of the matter-radiation fields and 
\begin{equation}\label{eq17}
t^{\lambda\mu}=k\left(4\Sigma^{bc\lambda}T_{bc}\,^{\mu}-g^{\lambda\mu}\Sigma^{bcd}T_{bcd}\right)\,.
\end{equation}
Field equations (\ref{eq16}) transform covariantly under coordinate transformations, and under global and local Lorentz transformations. By writing them as $ R_{a\mu} - \frac{1}{2} e_{a\mu} R = \frac{1}{2k} T_{a\mu} $, we can see that they are equivalent to Einstein's equations. Hence, all known solutions of GR are solutions to TEGR.

The advantage of the form (\ref{eq16}) is the possibility of defining a true energy-momentum tensor for the gravitational field. Given the antisymmetry of the superpotential in the last two indices, we have a null divergence for (\ref{eq17}), i.e.,
\begin{equation}\label{eq18}
\partial_{\lambda}\partial_{\nu}\left(e\,\Sigma^{a\lambda\nu}\right)=0\,.
\end{equation}
By separating the spatial and temporal components, and integrating them over an arbitrary volume $ V $, we obtain
\begin{equation}\label{eq19}
\frac{d}{dt}\int_{V}{\partial_{i}\left(e\,\Sigma^{a0i}\right)dV}=-\oint_{\partial V}{dS_{j}\partial_{\nu}\left(e\,\Sigma^{aj\nu}\right)}\,,
\end{equation}
where $ \partial V $ is the contour of $ V $. If we take $ \partial V \rightarrow \infty $ and the gravitational field goes to zero fast enough, the right-hand side of (\ref{eq19}) is zero. Hence, we have the conservation of the quantity
\begin{equation}\label{eq20}
P^{a}=4k\int_{V}{\partial_{i}\left(e\,\Sigma^{a0i}\right)dV}
\end{equation}
in time. Therefore, we have $ P^{a} $ as the total energy-momentum contained within the volume $ V $ of space and $ P^{(0)} $ as the total energy. 
Since quantity (\ref{eq17}) appears together with the energy-momentum tensor of matter, we identify it as the energy-momentum tensor of the gravitational field. This quantity is a true tensor, transforming covariantly under coordinate transformations and invariant under global Lorentz transformations. The energy-momentum 4-vector (\ref{eq20}) is invariant under spatial coordinate transformations, time reparametrizations, and global Lorentz transformations. It has been applied to several known solutions of Einstein's equations, yielding very satisfactory results consistent with others in the literature. For a recent review of some important solutions, see Ref. \cite{maluf2023tetrad}.

When the gravitational field does not decay sufficiently fast at the surface $\partial V$, i.e., when the right-hand side of (\ref{eq19}) does not vanish, there is a nonzero energy-momentum flux across that surface. Hence, for $ a = (0) $, we have
\begin{equation}\nonumber
    \frac{dP^{(0)}}{dt}=-\Phi_{g}^{(0)}-\Phi_{m}^{(0)}\,,
\end{equation}
where
\begin{equation}\nonumber
    \Phi_{g}^{(0)}=\oint_{\partial V}{dS_{j}e\, e^{(0)}\,_{\mu}t^{j\mu}}
\end{equation}
and
\begin{equation}\nonumber
    \Phi_{m}^{(0)}=\oint_{\partial V}{dS_{j}e\, e^{(0)}\,_{\mu}T^{j\mu}}
\end{equation}
are the gravitational and matter energy fluxes, respectively. This situation is typical of cosmological solutions. When evaluating the total energy at the apparent horizon, we find that the energy depends on the time coordinate, meaning that there is an energy-momentum flux across the apparent horizon \cite{da2018gravitational}.

\section{Szekeres spacetime}\label{sec3}
The spacetime considered by Szekeres consists of two sandwich gravitational waves propagating along null directions $ u $ and $ v $, respectively. A sandwich gravitational wave consists of a pulse of a plane wave, i.e., a spacetime region where we have a congruence of null geodesics without expansion, shear, and rotation, surrounded by two flat spacetime regions. The first flat spacetime region is ahead of the wave pulse, and the second behind, forming a sandwich.  As the duration of the pulse approaches zero, the sandwich wave reduces to an impulsive wave, in which the wavefront becomes an instantaneous disturbance, typically represented by a delta-function profile. Mathematically, following the notation of Ref. \cite{abbasi2021probing}, we have for sandwich wave 1 $(W_1)$, 
\begin{equation}
	\begin{cases}
		u<0: \text{flat spacetime}\nonumber\,,\\
		0\leq u \leq u_{0}: W_1\nonumber\,,\\
		u>u_{0}: \text{flat spacetime}\nonumber\,,
	\end{cases}
\end{equation}
where $ u = 0 $ and $ u = u_0 $ denote the ``borders" of $W_1$. Similarly, for wave 2 $(W_2)$, we have
\begin{equation}
	\begin{cases}
		v<0: \text{flat spacetime}\nonumber\,,\\
		0\leq v \leq v_{0}: W_2\nonumber\,,\\
		v>v_{0}: \text{flat spacetime}\nonumber\,,
	\end{cases}
\end{equation}
where $ v = 0 $ and $ v = v_0 $ denote the ``borders" of $W_2$. We shall assume that the waves are equal, i.e., have the same pulse profile.

The whole spacetime can be described by the line element \cite{abbasi2021probing}
\begin{equation}\label{eq24}
	ds^{2}=-2e^{-M}dudv + e^{V-U}dx^{2} + e^{-V-U}dy^{2}\,,
\end{equation}
where $M$, $U$, and $V$ are functions of $(u,v)$ in the general case. The coordinates $(x,y)$ span the 2-dimensional wavefront, and the null coordinates are related to the $(t,z)$ coordinates of Minkowski spacetime through
\begin{equation}
	u=(t-z)/\sqrt{2} ~\text{and}~~ v=(t+z)/\sqrt{2}.
\end{equation}
In $(t,x,y,z)$ coordinates, $W_1$ propagates along $+z$, and wave 2 along $-z$. In the double null coordinates $(u,v)$, the waves collide at an angle of 45 degrees, but in $(t,x,y,z)$ coordinates, they collide front to front.

To simplify the analysis, it is often advantageous to work with dimensionless quantities, as these provide a normalized framework that makes the results independent of specific unit systems. To achieve this normalization, we introduce the parameter $T$, which has the same dimensions as time. By dividing the null coordinates $u$  and $v$ by $T$, we redefine them as,
\begin{equation}
 \frac{u}{T} = \frac{t - z}{\sqrt{2} T}   ~\text{and}~~\frac{v}{T} = \frac{t + z}{\sqrt{2} T}.
\end{equation}
Now, $\frac{u}{T}$ and $\frac{v}{T}$ are dimensionless variables, representing the normalized null directions in the sandwich pp-wave spacetime. The parameter $T$ acts as a scaling factor, ensuring dimensional consistency across the solutions in different regions of the spacetime.

\begin{figure}[htbp]
	\centering
		\includegraphics[width=0.50\textwidth]{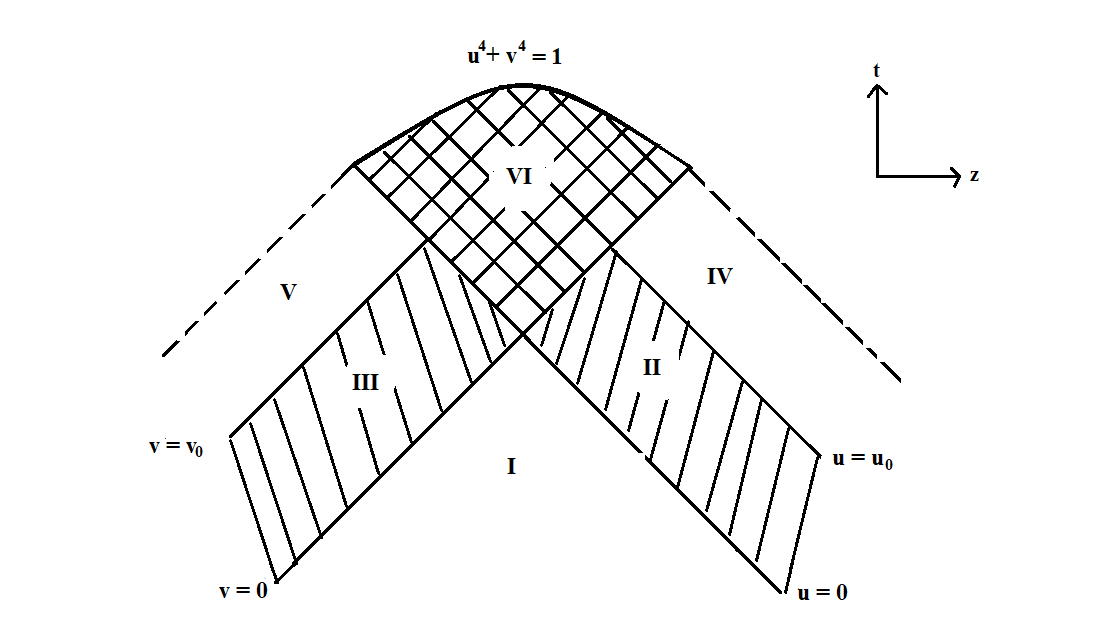}
	\caption{The Szekeres sandwich gravitational wave spacetime consists of six regions. Regions I, IV, and V are flat, while Regions II and III correspond to the approaching waves, and Region VI represents the post-collision region.}
	\label{fig0}
\end{figure}

The spacetime can be divided into six regions. The three flat regions are the spacetime sections behind the waves, the spacetime section after $W_1$, and the spacetime section after $W_2$; all of these are present before the collision. The two pp-wave spacetime sections are the sandwich waves of $W_1$ and $W_2$. The remaining spacetime section is the result of the collision. In each of these spacetime regions, the functions $M$, $U$, and $V$ have different expressions, but there is continuity on the borders between the regions \cite{szekeres1972colliding}.  The structure of the spacetime is illustrated in Figure \ref{fig0}.  Below, we summarize the form of the functions $M$, $U$, and $V$ in these six regions, and hence the metric in each one.

Region I is the flat region behind the waves described by
\begin{equation}
	I: u,v \leq 0\,,
\end{equation}
where the functions are $M = 0$, $U = 0$, and $V = 0$.

Region II is $W_1$ before the collision, described by
\begin{equation}
	II: v \leq 0\,, \qquad 0 \leq u \leq u_{0}\,,
\end{equation}
where the functions are
\begin{align}
II = \begin{cases}
U(u) &= -\ln{(1 - u^4/T^{4})}\,\,, \\
V(u) &= \sqrt{6} \arctan(u^2/T^{2})\,, \\
M(u) &= -\frac{1}{4} \ln{(1 - u^4/T^{4})}\,.
\end{cases}
\end{align}

Region III is  $W_2$ before the collision, described by
\begin{equation}
	III: u \leq 0\,, \qquad 0 \leq v \leq v_{0}\,,
\end{equation}
where the functions are
\begin{align}
III = \begin{cases}
U(v) &= -\ln{(1 - v^4/T^{4})}\,\,, \\
V(v) &= \sqrt{6} \arctan(v^2/T^{2})\,, \\
M(v) &= -\frac{1}{4} \ln{(1 - v^4/T^{4})}\,.
\end{cases}
\end{align}

Region IV is the flat region ahead of  $W_1$, described by
\begin{equation}
	IV: v \leq 0\,, \qquad u \geq u_{0}\,,
\end{equation}
where the functions are $M = 0$, $U = 0$, and $V = 0$.

Region V is the flat region ahead of $W_2$, described by
\begin{equation}
	V: u \leq 0\,, \qquad v \geq v_{0}\,,
\end{equation}
where the functions are $M = 0$, $U = 0$, and $V = 0$.

Region VI is the spacetime section after the collision, described by
\begin{equation}
	VI: u \geq u_{0}\,, \qquad v \geq v_{0}\,.
\end{equation}
The functions are
\begin{widetext}
   \begin{align}
VI = \begin{cases}
U(u,v) &= -\ln{(1 - u^{4}/T^{4} - v^4/T^{4})}\,, \\
V(u,v) &= \sqrt{6} \arctan \left[ u^2(1 - v^{4}/T^{4})^{-1/2} /T^{2}\right] + \sqrt{6} \arctan \left[ v^2(1 - u^{4}/T^{4})^{-1/2}/T^{2} \right] \,, \\
M(u,v) &= \frac{3}{4} \ln{[(1 - u^4/T^{4})(1 - v^{2}/T^{2})]} - \ln{(1 - u^{4}/T^{4} - v^4/T^{4})} \\
&+ 3 \arctan \left[ u^{2} v^{2} (1 - u^4/T^{4})^{-1/2} (1 - v^{4}/T^{4})^{-1/2}/T^{4} \right]\,.
\end{cases}
\end{align}  
\end{widetext}

We may note that there are singularities: in Region II at $u = T$, in Region III at $v = T$, and in Region VI at $u^{4} + v^{4} = T^{4}$. These singularities are present not only in the metric but also in the Weyl tensor components, as discussed by Szekeres himself \cite{szekeres1972colliding}. The Weyl tensor components are presented also in Ref. \cite{abbasi2021probing}, and the singularities are further discussed elsewhere \cite{aoki2023colliding}.

\section{Gravitational energy of Szekeres spacetime}\label{sec4}

The evaluation of the energy-momentum 4-vector begins with the establishment of a set of tetrads associated with the metric tensor. The definition of the energy-momentum is performed by foliating spacetime in hypersurfaces $t = \text{constant}$. Therefore, it is convenient to write the metric (\ref{eq24}) in Cartesian-like coordinates as
\begin{equation}\label{eq27}
	ds^2 = -e^{-M} dt^2 + e^{V-U} dx^2 + e^{-U-V} dy^2 + e^{-M} dz^2\,.
\end{equation}
There are an infinite number of tetrads associated with the metric (\ref{eq27}). However, some of them do not make the torsion tensor vanish in the Minkowski limit. Therefore, we require $e_{a}\,^{\mu} \rightarrow \delta_{a}^{\mu}$ in this limit.

One simple set of tetrads that satisfies the above requirement is
\begin{equation}\small
	e_{a\mu} = \text{diag}(-e^{-M/2}\,,\, e^{(V-U)/2}\,,\, e^{(-U-V)/2}\,,\, e^{-M/2})\,.\label{eq28}
\end{equation}
We will also need its inverse, given by
\begin{equation}\small
	e^{a\mu} = \text{diag}(e^{M/2}\,,\, e^{-(V-U)/2}\,,\, e^{-(-U-V)/2}\,,\, e^{M/2})\,,
\end{equation}
and its determinant is given by
\begin{equation}
	e = \det(e^{a}\,_{\mu}) = e^{-M-U}\,.
\end{equation}

The choice of tetrads plays a fundamental role in the determination of energy within TEGR. While an infinite number of tetrads satisfy equation (\ref{eq16}), different choices correspond to different observers and, consequently, different energy measurements. In this work, the tetrads (\ref{eq28}) were not selected merely for simplicity but rather due to their adaptation to a stationary, non-rotating observer in the considered spacetime. This ensures that the energy evaluation is not affected by additional motion-related effects, as it is shown in the next subsection. If an alternative tetrad set were chosen, the computed energy would differ, as it would be associated with a different observer. This does not imply a preferred frame but highlights the observer dependence of gravitational energy, similar to the situation in special relativity. The construction of a Fermi-Walker transported frame from an arbitrary tetrad field is possible in principle but not always straightforward \cite{maluf2008construction}, reinforcing the practical choice of analyzing the kinematic properties of a given tetrad rather than constructing one from first principles. Thus, our choice provides a clear and elegant interpretation of the energy measurement.

Before evaluating the energy, we must understand the observer associated with the tetrads (\ref{eq28}).

\subsection{Observer's inertial acceleration}

We may notice that, when $M = U = V = 0$, we have $e_{a}\,^{\mu} = diag(1, 1, 1, 1)$ as expected. Since $e_{(0)}\,^{\mu} = u^{\mu} = (u^{0}, 0, 0, 0)$, we can observe that the observer is stationary, as seen by another static observer at infinity (far from the waves' effects). However, in the absence of a global timelike Killing vector in the colliding wave regions, our definition of a ``static observer" does not rely on symmetries. In this frame, the observer's four-velocity remains aligned with the tetrad component $e_{(0)}{}^{\mu}$, ensuring $u^i = 0$.

In TEGR, the energy-momentum $P^a$ depends explicitly on the tetrad field $e_{a}{}^{\mu}$, which encodes both spacetime geometry and the observer's inertial properties through the acceleration tensor $\phi_{ab}$. The chosen tetrad, adapted to Cartesian-like coordinates, ensures that the observer's velocity follows $e_{(0)}{}^{\mu} = (e^{-M/2}, 0, 0, 0)$, reinforcing the frame-dependent nature of our approach. Importantly, while spatial coordinate transformations and time reparametrizations preserve the energy definition, nonlinear temporal redefinitions may alter the observer's velocity, leading to different energy measurements. To maintain consistency, we impose that a static observer at infinity satisfies $e_a{}^{\mu} = \delta_a^{\mu}$, ensuring a well-defined energy interpretation.

Additionally, Schwinger identified an incompleteness in the tetrad formulation of General Relativity regarding temporal evolution \cite{schwinger1963quantized}. To address this, he introduced the condition $e_{(0)}{}^i = 0$, known as the Schwinger gauge, which guarantees a well-defined evolution of the system. Our work adopts this gauge to ensure that the observer's velocity remains purely timelike.


The next step is the evaluation of the torsion tensor components (\ref{eq9}). The evaluation is rather simple for the diagonal tetrads (\ref{eq28}), and we list below the non-zero components of 
$T^{abc} = e^{b\mu} e^{c\nu} T^{a}\,_{\mu\nu}$ needed to evaluate the superpotential:
\begin{align}
	T^{(0)(0)(3)} &= \frac{e^{M/2}}{\sqrt{8}} (\partial_{u}M - \partial_{v}M)\,, \\
	T^{(1)(0)(1)} &= \frac{e^{M/2}}{\sqrt{8}} (\partial_{u}U+\partial_{v}U-\partial_{v}V   - \partial_{u}V)\,, \\
	T^{(1)(1)(3)} &= \frac{e^{M/2}}{\sqrt{8}} (\partial_{v}U + \partial_{u}V - \partial_{v}V - \partial_{u}U)\,, \\
	T^{(2)(0)(2)} &= \frac{e^{M/2}}{\sqrt{8}} (\partial_{u}U + \partial_{u}V+\partial_{v}U + \partial_{v}V)\,, \\
	T^{(2)(2)(3)} &= \frac{e^{M/2}}{\sqrt{8}} (\partial_{v}U + \partial_{v}V - \partial_{u}U - \partial_{u}V)\,, \\
	T^{(3)(0)(3)} &= \frac{e^{M/2}}{\sqrt{8}} (\partial_{u}M + \partial_{v}M)\,.             
\end{align}

With the torsion tensor components, we may evaluate the only non-zero component of the acceleration tensor as
\begin{align}\label{eq37}
	\phi_{z}(u,v) = \phi_{(0)(3)} = T_{(0)(0)(3)} &= \frac{1}{\sqrt{8}} e^{M/2} (\partial_{u}M - \partial_{v}M)\nonumber \\
	&= -\frac{1}{2} e^{M/2} \partial_{z}M\,.
\end{align}
Inertial acceleration (\ref{eq37}) keeps the observer in its stationary state $u^{i} = 0$.

For consistency, we observe that in spacetime Regions I, IV, and V, the inertial acceleration is zero, as expected, since no inertial acceleration is required to maintain a static observer in their kinematic state. In Region II, we have
\begin{equation}
	\phi_{ZII} = \frac{u^3}{ \sqrt{8} T^4 }\left(1-\frac{u^4}{T^4}\right)^{-9/8} \geq 0\,,
\end{equation}
for $0 \leq u \leq u_{0}< T$.
Since the observer's acceleration is positive along the $z$ axis, we conclude that the gravitational acceleration is negative along the $z$ axis.
In Region III, we have
\begin{equation}
	\phi_{ZIII} = - \frac{v^3}{ \sqrt{8} T^4 }\left(1-\frac{v^4}{T^4}\right)^{-9/8} \leq 0\,,
\end{equation}
for $0 \leq v \leq v_{0} < T$.
Since the observer's acceleration is negative along the $z$ axis, we conclude that the gravitational acceleration is positive along the $z$ axis.
We may conclude that both waves try to drag the observers contrary to their propagation direction, i.e., towards its origin.

It is interesting to note that the function $M(u)$ can be locally removed in Region II by defining a new $v$ coordinate as
\begin{equation}
	v' = \int e^{-M(u)}\,dv + \text{constant}\,.
\end{equation}
Therefore, the line element reads
\begin{equation}
	ds^{2} = -2\,du\,dv' + e^{V-U}\,dx^{2} + e^{-V-U}\,dy^{2}\,.
\end{equation}
Considering the above metric, the acceleration tensor vanishes, and we have a free-falling observer. In the case of Schwarzschild geometry, such an observer measures zero energy for the gravitational field \cite{maluf2007reference}, which is interpreted as a consequence of the Principle of Equivalence. However, in pp-wave spacetimes, this is not the case, as discussed in Ref. \cite{formiga2018energy}.

For Region VI, the expression is lengthy and depends on both $u$ and $v$. The signs will change as these coordinates are modified. However, there are two interesting results. They are
\begin{equation}
	\phi_{zVI}(0,0) = 0\,,
\end{equation}
and
\begin{equation}
	\lim_{(u^{4}+v^{4}) \to T^{4}} \phi_{zVI} = \pm\infty\,.
\end{equation}
Thus, the observer experiences no gravitational acceleration at the moment of the collision and requires infinite acceleration to maintain its stationary state close to the singularity.

\subsection{Integration of the energy density}

We may now proceed to evaluate the gravitational energy of the Szekeres spacetime as measured by a static observer. In order to compute the energy, we need the superpotential components. We do not need to evaluate all components, because only the component
\begin{equation}
	\Sigma^{(0)0j} = e_{(0)}{}^{0} e_{(j)}{}^{j} \Sigma^{(0)(0)(j)}
\end{equation}
will be relevant for the calculations. The diagonal form of the tetrads makes the calculations relatively simple. By noting that
\begin{equation}\small
	\Sigma^{(0)(0)(j)} = \frac{1}{2} T^{(0)(0)(j)} - \frac{1}{4} \Sigma^{(j)(0)(0)} - \frac{1}{2} \eta^{(0)(0)} T^{a}\,_{a}\,^{(j)},
\end{equation}
we see that the only non-vanishing component is
\begin{align}
	\Sigma^{(0)(0)(3)} &= \frac{1}{2} \left( T^{(1)(1)(3)} + T^{(2)(2)(3)} \right) \nonumber\\
	&= \frac{e^{M/2}}{\sqrt{8}} \left( \partial_{v}U - \partial_{u}U \right),
\end{align}
with $\Sigma^{(0)(0)(1)} = 0 = \Sigma^{(0)(0)(2)}$.
Thus, we have
\begin{equation}
	\Sigma^{(0)03} = \frac{e^{3M/2}}{\sqrt{8}} \left( \partial_{v}U - \partial_{u}U \right),
\end{equation}
and the energy density is
\begin{align}
	dP^{(0)} &= 4k \, \partial_{z} \left( e \Sigma^{(0)03} \right)\nonumber \\
	&= 4k \, \partial_{z} \left[ \frac{e^{M/2-U}}{\sqrt{8}} \left( \partial_{v}U - \partial_{u}U \right) \right].
\end{align}

The total energy can be obtained by integrating the energy density over a volume $ V $, i.e., 
\begin{equation}
	P^{(0)} = 4k \int_{V} d^3x \, \partial_z \left[ \frac{e^{M/2-U}}{\sqrt{8}} \left( \partial_v U - \partial_u U \right) \right] \,.\label{eq50}
\end{equation}
The volume can range from Regions I to VI or any section of them, as the energy definition is local.

After choosing the volume $ V $, integral (\ref{eq50}) can be solved to obtain an expression for the energy contained within that volume. While possible, it is easier to work with a surface integral. By applying the Divergence Theorem to (\ref{eq50}), we obtain the same energy given by the surface integral
\begin{equation}
	P^{(0)} = 4k \oint_{\partial V} dS_z \, \left[ \frac{e^{M/2-U}}{\sqrt{8}} \left( \partial_v U - \partial_u U \right) \right] \,,\label{eq51}
\end{equation}
where $ \partial V $ is the boundary of $ V $.

Given the symmetry of the problem, we consider a box with sides of size $ L $ in the $ x $- and $ y $-directions and length $ z_2 - z_1 $, with the top and bottom of the box given by the surfaces $ z = z_2 = \text{constant} $ and $ z = z_1 = \text{constant} $. Hence, we have
\begin{align}
	E = P^{(0)} &= \sqrt{2} k \int_0^L dx \int_0^L dy \, \left[ e^{M/2-U} \left( \partial_v U - \partial_u U \right) \Big|_{z = z_2} \right. \nonumber\\
	& \quad \left. - e^{M/2-U} \left( \partial_v U - \partial_u U \right) \Big|_{z = z_1} \right]\nonumber\\
	&=\epsilon(z_{2})-\epsilon(z_{1}) \label{eq52}\,,
\end{align}
where we call $\epsilon$ ``energy density". this quantity is not the energy density but is a function of $(t,z)$ that allows us to analyze the energy distribution along the plane $u, v$.
If we extend the box to the entire spacetime, i.e., as $ L \to \infty $, we obtain infinite energy. This is expected since we are considering a plane wave that extends throughout the whole 2-dimensional perpendicular space spanned by $ (x, y) $. In a real wave, the plane front approximation can only be made over a finite region of the wavefront. Therefore, we regard the energy per square unit of area $ P^{(0)}/L^2 $ as the physical quantity of interest.

For consistency, we note that the energy is zero in Regions I, III, and V. 

For Region II, we have the energy density as a function of $u$ given by
\begin{equation}\label{eq55}
	\epsilon_{II}(u) = -\frac{L^2 u^3}{ \sqrt{8} \pi  T^4 \sqrt[8]{1-\frac{u^4}{T^4}}} \,.
\end{equation}
For Region III, but with $ v $, we get
\begin{equation}\label{eq56}
	\epsilon_{III}(v) = \frac{L^2 v^3}{ \sqrt{8} \pi  T^4 \sqrt[8]{1-\frac{v^4}{T^4}}} \,.
\end{equation}

Thus, the observer measures a negative energy density for  $W_1$ and a positive energy density for $W_2$, both of the same magnitude. The distinct signs arise because the integration surface is oriented along the positive $ z $-direction, i.e., $ z_2 > z_1 $. Consequently, the observer sees wave 1 pass through $ z_1 $ first, then $ z_2 $, while  $W_2$ passes through $ z_2 $ first, then $ z_1 $. If we were to reverse the orientation of the surface, the signs of the energies would be reversed. There is no inconsistency, as we are dealing with equal waves propagating in opposite directions.

\begin{figure}[htbp]
	\centering
		\includegraphics[width=0.40\textwidth]{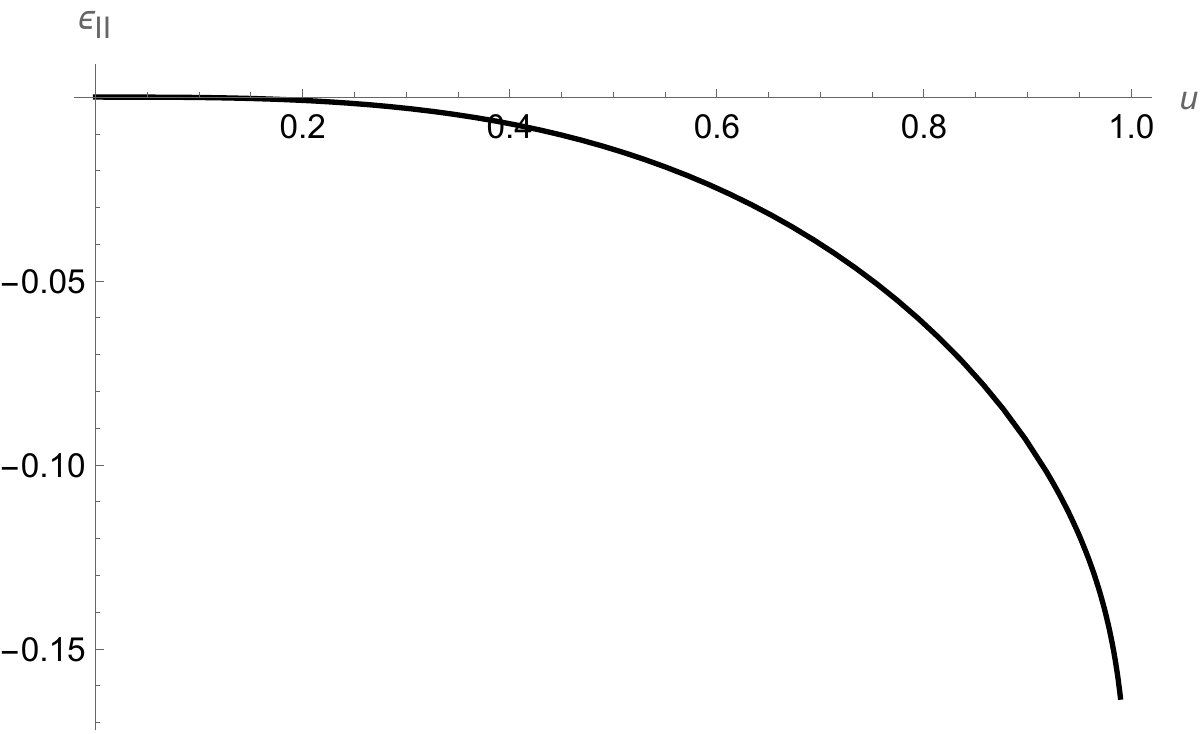}
	\caption{Energy density (\ref{eq55}) of $W_1$ as a function of $ u $.}
	\label{fig1}
\end{figure}

\begin{figure}[htbp]
	\centering
		\includegraphics[width=0.40\textwidth]{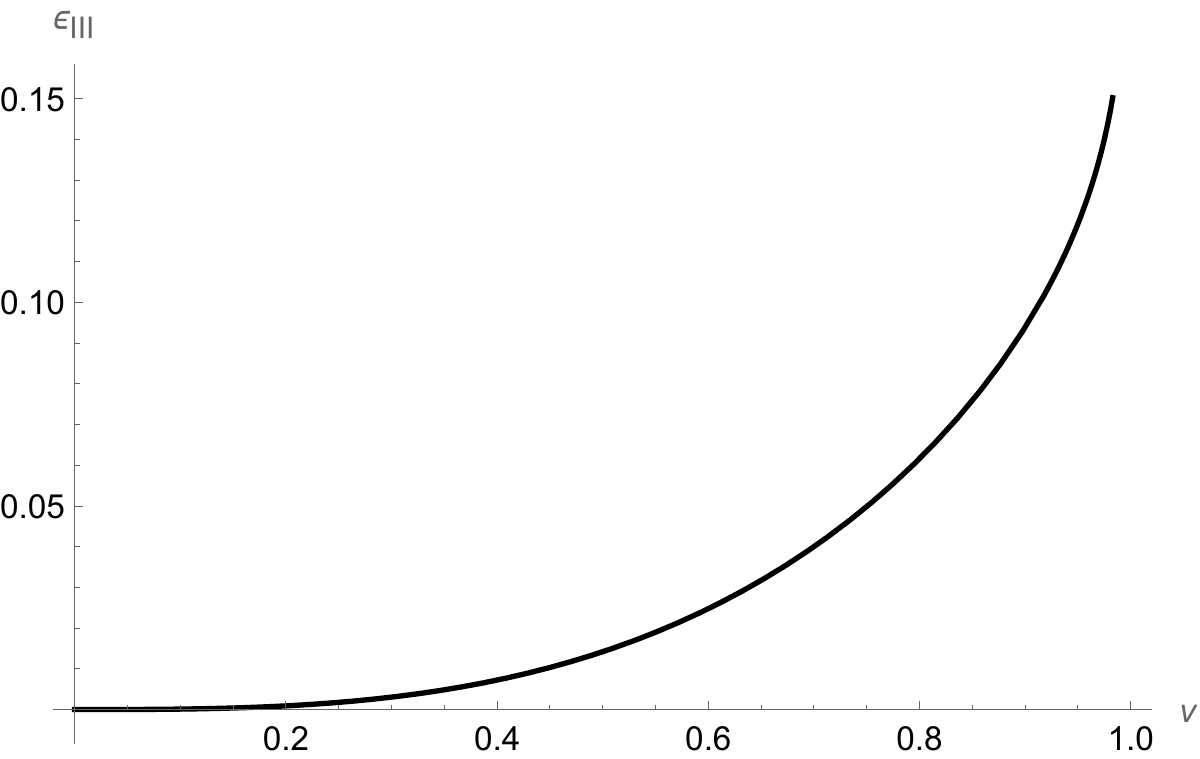}
	\caption{Energy density (\ref{eq56}) of  $W_2$ as a function of $ v $.}
	\label{fig2}
\end{figure}

The results are displayed in Figures \ref{fig1} and \ref{fig2}, where we have used $ L = 1 =T$ and shall maintain this choice throughout the rest of the article.
Note that we observe a variation in the energy flux, as we are considering a static integration surface where the observer field measures the wave passing through it. To maintain a constant energy value, we would need to consider an integration surface that moves with the wave, as it would always contain the wave within it. We will explore this analysis in the next section.

\begin{figure}[htbp]
	\centering
		\includegraphics[width=0.40\textwidth]{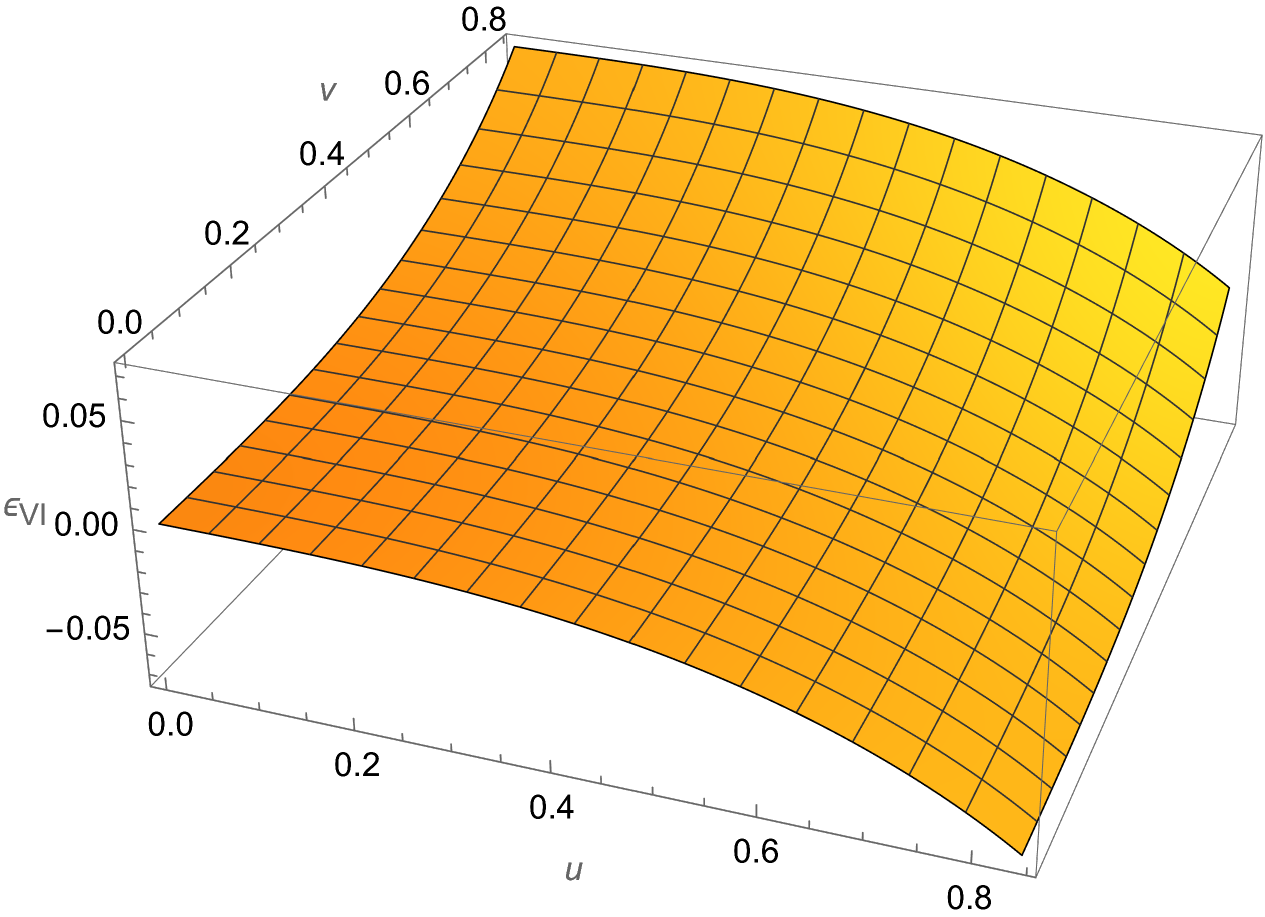}
	\caption{Gravitational energy density (\ref{eq56}) of the resulting spacetime as a function of $ u $ and $ v $.}
	\label{fig3}
\end{figure}

In Region VI, we cannot express the energy density as a function of $ u $ or $ v $ alone. Hence, we have a function of $(u,v)$ given by
\begin{align}
\epsilon_{VI} &= -\frac{L^2}{ \sqrt{8} \pi T^{4}}  \frac{ \left(u^3 - v^3\right) \left[\left(1-u^4/T^{4} \right)\left(1-v^{4}/T^{4}\right)\right]^{3/8} }{ \sqrt{1-u^4/T^{4} - v^4/T^{4} }} \nonumber \\
&\quad \cdot e^{\frac{3}{2} \tanh^{-1}\left( \frac{u^2 v^2/T^{4}}{\sqrt{1 - u^4/T^{4}} \sqrt{1 - v^4/T^{4}}} \right)}
\end{align}
as the energy density along the $ (u, v) $-plane. Hence, we can plot a surface representing the energy density distribution along the $ (u, v) $-plane, as shown in Figure \ref{fig3}. We observe regions of both positive and negative energy density.

\begin{figure}[htbp]
	\centering
		\includegraphics[width=0.40\textwidth]{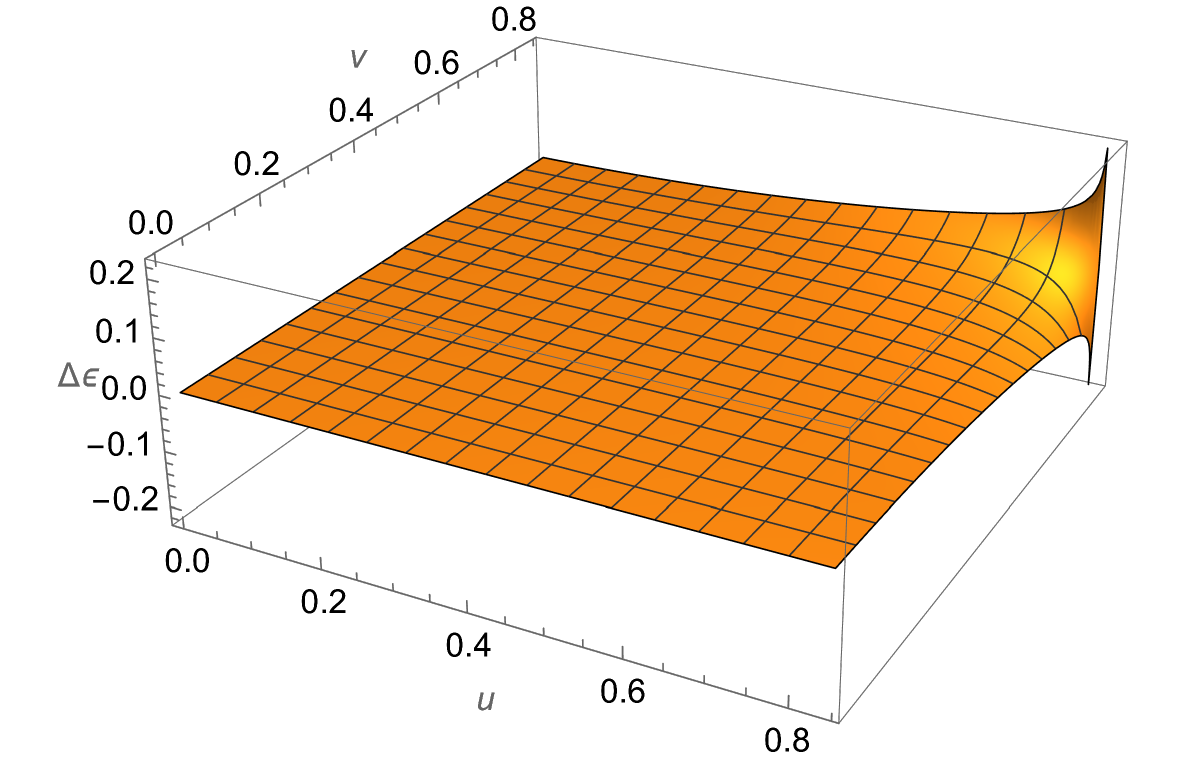}
	\caption{Energy density difference $ \Delta \epsilon = \epsilon_{VI} - (\epsilon_{II} + \epsilon_{III}) $ as a function of $ u $ and $ v $.}
	\label{fig4}
\end{figure}

Interestingly, if we compute the sum of the energy densities of $W_1$ and $W_2$ in the absence of the collision, i.e., the resulting energy from a linear interaction and superposition, and subtract this from the energy density of the resulting spacetime after the interaction, we observe an approximately zero energy region that covers almost the entire region (from the linear interaction), except near $ u^4 + v^4 = 1 $, i.e., near the singularity. The behavior is displayed in Figure \ref{fig4}.
This suggests the possibility of energy creation from the collision. We investigate this topic in the next section.

\section{Energy creation in the collision}\label{sec5}

In this section, we proceed to evaluate the gravitational energy contained in the same box of space considered within each region. We choose the integration volume to be the spatial section of spacetime associated with Regions II, III, and VI. Hence, we need to transform the boundaries from $ (u, v) $ coordinates to $ (t, z) $ coordinates.

For Region II, corresponding to wave 1, we have
\begin{equation}
	II: t - \sqrt{2} u_0   \leq z \leq t \,.
\end{equation}
To compute the energy of the entire wave, we choose moving surfaces that follow the anterior and posterior wavefronts. Thus, the posterior surface of the integration box is $ z_1 = t - \sqrt{2} u_0 $, and the anterior surface is $ z_2 = t $. In this configuration, we are always evaluating the energy of the whole wave since the integration volume follows the wave trajectory. If we choose a fixed surface, we would observe a time-dependent energy (an energy flux), as the wave passes through the surface, increasing the energy inside the surface and then lowering it.

Using equation (\ref{eq52}), we transform from $ u, v $ coordinates to the $ t $-coordinate and obtain
\begin{align}
	E_{II} &= \frac{1}{8 \pi } L^2 \Big( e^{M/2 - U} \partial_z U \Big) \Big|_t^{t - \sqrt{2} u_0} \nonumber \\
	&= \frac{L^2 u_0^3}{ \sqrt{8} \pi T^{4} \sqrt[8]{1 - u_0^4/T^{4}}} \,. \label{eq61}
\end{align}
As expected, the energy is constant for the wave when using the moving surface. By choosing the value of width $ u_0 $, we can evaluate the energy per unit of transversal area for  $W_1$.

Similarly, for Region III, corresponding to $W_2$, we have
\begin{equation}
	III: -t \leq z \leq \sqrt{2} v_0 - t \,,
\end{equation}
and consequently,
\begin{align}
	E_{III} &= \frac{1}{8 \pi} L^2 \Big( -e^{M/2 - U} \partial_z U \Big) \Big|_{-t}^{\sqrt{2} v_0 - t} \nonumber \\
	&= \frac{L^2 v_0^3}{ \sqrt{8} \pi T^{4}  \sqrt[8]{1 - v_0^4/T^{4}}} \,. \label{eq63}
\end{align}
If $ u_0 = v_0 $, we note that the energy is the same for both waves, as expected. As discussed in the previous section, the use of null coordinates $ u $ and $ v $ can be misleading, as they may suggest that the waves have different energy signs. 

The waves considered here are the $ + $ polarization waves studied by Maluf and Ulhoa in Brinkmann coordinates \cite{maluf2008energy} but for a general pulse. In that work, they obtained a negative energy value, but here we are considering a different set of observers. The observers adapted to the tetrads in equation (\ref{eq28}) do not rotate, as the components $ \phi_{(i)(j)} $ of the acceleration are zero (for a recent discussion on the frame rotation, see Ref. \cite{formiga2022teleparallel}).

Now we turn our attention to Region VI, where
\begin{equation}
	VI: t - \sqrt{2} u_0  \leq z \leq \sqrt{2} v_0  - t \,.
\end{equation}
We can determine the initial time of the collision by analyzing the boundaries between Regions II, III, and Region VI. The initial time is given by
\begin{equation}
	t_0 = \frac{u_0 + v_0}{\sqrt{2}}  \,.
\end{equation}
The energy in Region VI is given by
\begin{widetext}
\begin{align}
E_{VI}(t) &= \frac{L^2}{4 \pi T^4} \Bigg[
\frac{\left\{ \left[ t - \sqrt{2} u_0 \right] \left[ 3t^2 + \left( t - \sqrt{2} u_0 \right)^2 \right] \right.}{
\sqrt{1-\frac{\left( \sqrt{2} u_0 - 2t \right)^4}{4 T^4} - \frac{u_0^4}{T^4} }} \notag \\
&\quad \times \left. \left( \left( 1 - \frac{u_0^4}{T^4} \right) 
\left( 1 - \frac{\left( \sqrt{2} u_0 - 2t \right)^4}{4 T^4} \right) \right)^{3/8} \right\}  \exp \left\{ \frac{3}{2} \tanh^{-1} \left[ 
\frac{u_0^2 \left( \sqrt{2} u_0 - 2t \right)^2}{
2 T^4 \sqrt{1 - \frac{u_0^4}{T^4}} 
\sqrt{1 - \frac{\left( \sqrt{2} u_0 - 2t \right)^4}{4 T^4}}} \right] \right\} \notag \\
&\quad - \frac{\left\{ \left[ \sqrt{2} v_0 - t \right] \left[ 2t^2 - \sqrt{2} t v_0 + v_0^2 \right] \right.}{
\sqrt{-\frac{\left( \sqrt{2} v_0 - 2t \right)^4}{4 T^4} - \frac{v_0^4}{T^4} + 1}} \notag \\
&\quad \times \left. \left( \left( 1 - \frac{v_0^4}{T^4} \right) 
\left( 1 - \frac{\left( \sqrt{2} v_0 - 2t \right)^4}{4 T^4} \right) \right)^{3/8} \right\}  \exp \left\{ \frac{3}{2} \tanh^{-1} \left[ 
\frac{v_0^2 \left( \sqrt{2} v_0 - 2t \right)^2}{
2 T^4 \sqrt{1 - \frac{v_0^4}{T^4}} 
\sqrt{1 - \frac{\left( \sqrt{2} v_0 - 2t \right)^4}{4 T^4}}} \right] \right\}
\Bigg]\label{eq66}.
\end{align}
\end{widetext}

 Aoki \cite{aoki2023colliding} introduces an alternative definition for gravitational energy by considering singularities in spacetime. These singularities lead to non-zero contributions from the energy-momentum tensor (EMT), which allows the calculation of total matter-energy. This energy is not conserved during the collision of the waves, but a generalized ``gravitational charge" can be defined, which remains conserved before the collision and its value is zero.
 
\begin{figure}[htbp]
	\centering
		\includegraphics[width=0.40\textwidth]{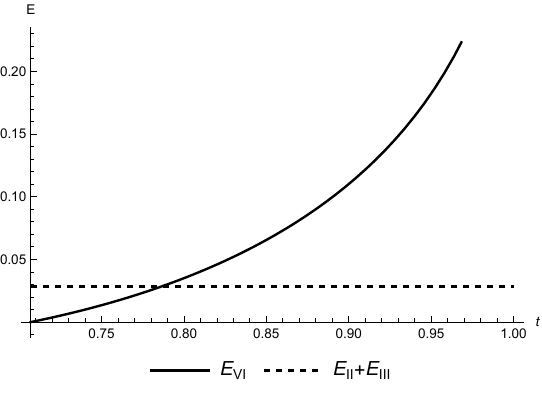}
	\caption{Comparison between the energy of Region VI and the combined energy of Regions II and III for $u_{0}=0.5=v_{0}$.}
	\label{fig5}
\end{figure}

Equation (\ref{eq66}) partially corroborates Aoki’s result because
\begin{equation}
	E_{VI}(t_0) = 0 \,,
\end{equation}
i.e., at the time of collision, the energy is zero. Aoki interprets this as the annihilation of the two waves. However, after the collision, we observe the energy to increase, surpassing the energy of the combined incoming waves, as shown in Figure \ref{fig5}.

This behavior is particularly interesting, as the energy vanishes after the collision, similar to a contraction, and then increases, akin to an expansion.
From Figure \ref{fig5}, we may infer that the apparent energy creation in Region VI arises from the nonlinear self-interaction of gravitational waves. This interaction restructures the torsional geometry of spacetime, leading to an emergent energy density not present in the incident waves. While local energy increases, global conservation may still hold when accounting for energy fluxes across the entire hypersurface. Alternatively, this process reveals the need to generalize energy conservation in spacetimes with strong-field interactions and nontrivial torsion structure.

\section{Conclusions}\label{conc}

In this article, we evaluated the gravitational energy of two colliding sandwich pp-waves described through Szekeres spacetime. We considered two identical sandwich $+$ polarization pp-waves moving in opposite directions and the resulting spacetime after the collision occurring after $ t = \frac{u_0 + v_0}{\sqrt{2}} $, where $ u_0 $ and $ v_0 $ are the widths of the waves. For the definition of gravitational energy, we used the conserved quantity arising from the vanishing divergence in the field equations of TEGR, which is a geometrical description of gravity that yields the same field equations as GR but allows for the definition of a consistent energy-momentum 4-vector. The conserved quantities of TEGR can also be obtained from the Hamiltonian formulation of the theory in vacuum, with the constraints satisfying the algebra of the Poincar\'e group \cite{maluf2013teleparallel}.

The energy here obtained is the energy that is expected to be measured by a static observer field that uses inertial acceleration (\ref{eq37}) to maintain its kinematic state. The triad $ e_{(i)}\,^{\mu} $ does not rotate, as is the case for the observer considered in Ref. \cite{maluf2008energy}, hence we expect distinct energy, since different non-inertial observers measure distinct values for the energy, as expected. For real measurements, one must conform with non-inertial frames, since they are the only ones available, as discussed by Formiga \cite{formiga2022teleparallel}. 

Interestingly, we may evaluate the energy of the wave contained within a region of spacetime. By choosing $ 10u_0 = L^2 = 1 \, m $, we obtain $ E = 1.125 \cdot 10^{-4} \, m = 1.125 \cdot 10^{-4} c^4 / G \, J = 1.36 \cdot 10^{40} \, J $, i.e., approximately $ 2.5\% $ of Earth's rest mass. It has a very large energy density and probably makes the non-linear effects of GR very appealing.

We obtained not only the gravitational energy of the colliding waves but also the analytical expression (\ref{eq61}) for the energy of the sandwich pp-wave. To our knowledge, such an expression has never been obtained before, as previous analyses required numerical integration of the energy density.
The energy was obtained using expression (\ref{eq20}), which originates from a balance equation. The gravitational energy is distributed along the entire wavefront, and since the latter extends to infinity, there is a nonzero energy flux at infinity. However, as a plane gravitational wave is an idealized approximation of a real wave, this behavior does not translate directly to a physical gravitational wave, whose wavefront would not extend infinitely. Moreover, if we were to choose an integration surface where energy varies over time (such as a fixed surface), we would observe an energy flux, as the energy would not be conserved within that surface.

Furthermore, the observer dependence of the energy raises an interesting discussion regarding the choice of tetrads. As shown in our analysis, different tetrad choices correspond to distinct observers, each measuring a different energy value. In particular, a change in the observer's frame may prevent the energy from vanishing after the collision, as it would incorporate additional energy contributions related to the motion of the new observer. However, while the numerical value of the energy depends on the chosen frame, the qualitative behavior remains unchanged, with energy first decreasing and then increasing after the collision. This suggests that the observed energy dynamics are not merely a frame effect but rather an intrinsic feature of the spacetime.

Our result shows that, immediately after the collision, the waves cancel out, since the energy drops to zero. This is a phenomenon not observed in the interaction of electromagnetic waves, but it is expected that non-linear effects of Einstein's field equation generate unfamiliar effects. After the collision, the energy of the resulting spacetime increases with time, surpassing the combined energy of the two incoming waves. This presents a very interesting prospect since the resulting spacetime of two colliding pp-waves is something else. The astrophysical consequences cannot be mitigated, since collisions between pp-waves (at least for those with $ + $ polarization) can annihilate the waves and create energy (\textit{ex nihilo?}). The prospect of annihilation is that the universe may not have many gravitational waves ``available" for detection, as it depends on a free path between us and the source. Although the waves vanish, they generate energy, which may increase the overall energy of the universe.

Although the considered case of Szekeres spacetime is very artificial (plane front, same wave, fixed polarization), it is strange that this feature is exclusive to this highly symmetric solution, because this high symmetry can be close to real gravitational waves. We believe that the comprehension requires further investigation of the spacetime after the collision, with the inertial accelerations computed and the geodesic motion analyzed. This may be pursued elsewhere.

It is also worthwhile to investigate the energy dynamics by considering two unequal-strength sandwich pp-waves using the same procedure. Unlike the case of equal-strength waves, the two waves will not cancel each other out. Consequently, the energy in Regions II and III would differ, and Region VI would no longer exhibit symmetry along $z=0$. This asymmetry implies an increase in energy, and it would be interesting to quantify the extent of this increase in comparison to the symmetric case (as done in this study).

Additionally, it would be worth exploring whether the energy expression for the unequal-strength waves remains superposed, as it does in the symmetric case. Furthermore, by taking the limit as the width of the sandwich tends to zero, the configuration transitions to an impulsive pp-wave \cite{Griffiths2016}. A potential direction for future research would be to examine how the energy in the interaction region relates to the case of an impulsive wave collision. This analysis could provide deeper insights into the relationship between sandwich waves and their impulsive counterparts.



\begin{thebibliography}{10}

\bibitem{bondi1957plane}
H.~Bondi.
\newblock Plane gravitational waves in general relativity.
\newblock {\em Nature}, 179(4569):1072--1073, 1957.

\bibitem{bondi1959gravitational}
H~Bondi, F.~A.~E. Pirani, and I.~Robinson.
\newblock Gravitational waves in general relativity iii. exact plane waves.
\newblock {\em Proceedings of the Royal Society of London. Series A. Mathematical and Physical Sciences}, 251(1267):519--533, 1959.

\bibitem{abid2024identification}
S.~Abid, K.~Atifi, El-H. Essoufi, and A.~Zafrar.
\newblock Identification of source term in a nonlinear degenerate parabolic equation with memory.
\newblock {\em Applications of Mathematics}, 69(2):209--232, 2024.

\bibitem{favata2010gravitational}
M.~Favata.
\newblock The gravitational-wave memory effect.
\newblock {\em Classical and Quantum Gravity}, 27(8):084036, 2010.

\bibitem{zhang2017memory}
P-M. Zhang, C.~Duval, G.~W. Gibbons, and P.~A. Horvathy.
\newblock The memory effect for plane gravitational waves.
\newblock {\em Physics Letters B}, 772:743--746, 2017.

\bibitem{chakraborty2022simple}
I.~Chakraborty and S.~Kar.
\newblock A simple analytic example of the gravitational wave memory effect.
\newblock {\em The European Physical Journal Plus}, 137(4):1--14, 2022.

\bibitem{zhang2018memory}
P.~M. Zhang, C.~Duval, and P.~A. Horvathy.
\newblock Memory effect for impulsive gravitational waves.
\newblock {\em Classical and Quantum Gravity}, 35(6):065011, 2018.

\bibitem{datta2024memory}
S.~Datta and S.~Guha.
\newblock Memory effect of gravitational wave pulses in pp-wave spacetimes.
\newblock {\em Physica Scripta}, 99(7):075023, 2024.

\bibitem{hait2024frequency}
A.~Hait, S.~Mohanty, and S.~Prakash.
\newblock Frequency space derivation of linear and nonlinear memory gravitational wave signals from eccentric binary orbits.
\newblock {\em Physical Review D}, 109(8):084037, 2024.


\bibitem{cvetkovic2022memory}
B.~Cvetkovi{\'c} and D.~Simi{\'c}.
\newblock Memory effect of the pp waves with torsion.
\newblock {\em The European Physical Journal C}, 82(2):127, 2022.

\bibitem{sun2023detecting}
S.~Sun, C.~Shi, J.-D. Zhang, and J.~Mei.
\newblock Detecting the gravitational wave memory effect with tianqin.
\newblock {\em Physical Review D}, 107(4):044023, 2023.

\bibitem{bhattacharya2024gravitational}
S.~Bhattacharya, D.~Bose, I.~Chakraborty, A.~Hait, and S.~Mohanty.  
\newblock Gravitational memory signal from neutrino self-interactions in supernova.  
\newblock {\em Physical Review D}, 110(6):L061501, 2024.  


\bibitem{zhang2018velocity}
P.~M. Zhang, C.~Duval, G.~W. Gibbons, and P.~A. Horvathy.
\newblock Velocity memory effect for polarized gravitational waves.
\newblock {\em Journal of Cosmology and Astroparticle Physics}, 2018(05):030, 2018.

\bibitem{divakarla2021first}
A.~K. Divakarla and B.~F. Whiting.
\newblock First-order velocity memory effect from compact binary coalescing sources.
\newblock {\em Physical Review D}, 104(6):064001, 2021.

\bibitem{zhang2024displacement}
P.~M. Zhang and P.~A. Horvathy.
\newblock Displacement within velocity effect in gravitational wave memory.
\newblock {\em Annals of Physics}, 470:169784, 2024.

\bibitem{zhang2024gravitational}
P.~M. Zhang, Q.~L. Zhao, M.~Elbistan, and P.~A. Horvathy.
\newblock Gravitational wave memory: further examples.
\newblock {\em arXiv preprint arXiv:2412.02705}, 2024.

\bibitem{maluf2018plane}
J.~W. Maluf, J.~F. Rocha-Neto, S.~C. Ulhoa, and F.~L. Carneiro.
\newblock Plane gravitational waves, the kinetic energy of free particles and the memory effect.
\newblock {\em Gravitation and Cosmology}, 24:261--266, 2018.

\bibitem{szekeres1970colliding}
P.~Szekeres.
\newblock Colliding gravitational waves.
\newblock {\em Nature}, 228(5277):1183--1184, 1970.

\bibitem{szekeres1972colliding}
P.~Szekeres.
\newblock Colliding plane gravitational waves.
\newblock {\em Journal of Mathematical Physics}, 13(3):286--294, 1972.

\bibitem{moller1964conservation}
C.~Moller.
\newblock Conservation laws in the tetrad theory of gravitation.
\newblock In {\em Proceedings of the Conference on Theory of Gravitation, Warszawa and Jablonna, 1962}. Gauthier-Villars, PWN-Polish Scientific Publishers, 1964.

\bibitem{maluf1995localization}
J.~W. Maluf.
\newblock Localization of energy in general relativity.
\newblock {\em Journal of Mathematical Physics}, 36(8):4242--4247, 1995.

\bibitem{formiga2023gravitational}
J.~B. Formiga and J.~A.~C. Duarte.
\newblock Gravitational energy problem and the energy of photons.
\newblock {\em Physical Review D}, 108(4):044043, 2023.

\bibitem{maluf2008energy}
J.~W. Maluf and S.~C. Ulhoa.
\newblock The energy-momentum of plane-fronted gravitational waves in the teleparallel equivalent of gr.
\newblock {\em Physical Review D---Particles, Fields, Gravitation, and Cosmology}, 78(4):047502, 2008.

\bibitem{mashhoon2002length}
B.~Mashhoon and U.~Muench.
\newblock Length measurement in accelerated systems.
\newblock {\em Annalen der Physik}, 514(7):532--547, 2002.

\bibitem{mashhoon2003vacuum}
B.~Mashhoon.
\newblock Vacuum electrodynamics of accelerated systems: Nonlocal maxwell's equations.
\newblock {\em Annalen der Physik}, 515(10):586--598, 2003.

\bibitem{maluf2013teleparallel}
J.~W. Maluf.
\newblock The teleparallel equivalent of general relativity.
\newblock {\em Annalen der Physik}, 525(5):339--357, 2013.

\bibitem{maluf2023tetrad}
J.~W. Maluf, F.~L. Carneiro, S.~C. Ulhoa, and J.~F. da~Rocha-Neto.
\newblock Tetrad fields, reference frames, and the gravitational energy--momentum in the teleparallel equivalent of general relativity.
\newblock {\em Annalen der Physik}, 535(12):2300241, 2023.

\bibitem{da2018gravitational}
J.~F.~da Rocha-Neto and B.~R.~Morais.  
\newblock Gravitational pressure, apparent horizon and thermodynamics of FLRW universe in the teleparallel gravity.  
\newblock {\em General Relativity and Gravitation}, 50:1--16, 2018.  

\bibitem{abbasi2021probing}
K.~Q. Abbasi, I.~Hussain, and A.~Qadir.
\newblock Probing szekeres’ colliding sandwich gravitational waves.
\newblock {\em The European Physical Journal Plus}, 136(5):565, 2021.

\bibitem{aoki2023colliding}
S.~Aoki.  
\newblock Colliding gravitational waves and singularities.  
\newblock {\em International Journal of Modern Physics A}, 38(21):2350120, 2023.  

\bibitem{maluf2008construction}
J.~W. Maluf and F.~F. Faria.  
\newblock On the construction of Fermi-Walker transported frames.  
\newblock {\em Annalen der Physik}, 520(5):326--335, 2008.  

\bibitem{schwinger1963quantized}
J.~Schwinger.
\newblock Quantized gravitational field.
\newblock {\em Physical Review}, 130(3):1253, 1963.

\bibitem{maluf2007reference}
J.~W. Maluf, F.~F. Faria, and S.~C. Ulhoa.
\newblock On reference frames in spacetime and gravitational energy in freely falling frames.
\newblock {\em Classical and Quantum Gravity}, 24(10):2743, 2007.

\bibitem{formiga2018energy}
J.~B. Formiga.
\newblock The energy-momentum tensor of gravitational waves, wyman spacetime, and freely falling observers.
\newblock {\em Annalen der Physik}, 530(12):1800320, 2018.

\bibitem{formiga2022teleparallel}
J.~B. Formiga.
\newblock On the teleparallel frame problem.
\newblock {\em Modern Physics Letters A}, 37(33n34):2250222, 2022.

\bibitem{Griffiths2016}
J.~B. Griffiths and J.~Podolsk{\'y}.
\newblock Exact solutions of einstein's field equations.
\newblock {\em Cambridge University Press}, 2016.

\end{thebibliography}

\end{document}